# Black-silicon ultraviolet photodiodes achieve external quantum efficiency above 130%

*M. Garin,*[1,2,3†] *J. Heinonen,*[1,4] *L. Werner*[5], *T. P. Pasanen,*[1] *V. Vähänissi,*[1] *A. Haarahiltunen,*[4] *M. Juntunen,*[4] *H. Savin,*[1*]

[1] Department of Electronics and Nanoengineering, Aalto University, Tietotie 3, 02150 Espoo, Finland
[2] Department of Engineering, Universitat de Vic – Universitat Central de Catalunya, c/ de la Laura 13, 08500 Vic, Spain.
[3] Univ. Politècnica de Catalunya, Gran Capità s/n, 08034 Barcelona, Spain
[4] ElFys inc, Tekniikantie 12, 02150 Espoo, Finland
[5] Physikalisch-Technische Bundesanstalt, Abbestr. 2-12, 10587 Berlin, Germany

†moises.garin@uvic.cat
* hele.savin@aalto.fi

*At present, ultraviolet sensors are utilized in numerous fields ranging from various spectroscopy applications via biotechnical innovations to industrial process control. Despite of this, the performance of current UV sensors is surprisingly poor. Here, we break the theoretical one photon – one electron barrier and demonstrate a device with a certified external quantum efficiency (EQE) above 130% in UV range without external amplification. The record high performance is obtained using a nanostructured silicon photodiode with self-induced junction. We show that the high efficiency is based on effective utilization of multiple carrier generation by impact ionization taking place in the nanostructures. While the results can readily have a significant impact on the UV-sensor industry, the underlying technological concept can be applied to other semiconductor materials, thereby extending above unity response to longer wavelengths and offering new perspectives for improving efficiencies beyond the Shockley-Queisser limit.*

Ultraviolet (UV) sensors are currently being utilized in a wide range of applications, including spectroscopy, imaging, flame detection, water purification and biotechnology—just to name a few. [1–6] Furthermore, an annual market growth rate of about 30% is expected [7]. Therefore, it is quite surprising that the semiconductor sensors available in the market suffer from relatively poor UV response, the best sensitivities falling well below 80% at 200-300 nm [8, 9]. All UV applications would greatly benefit from better response, raising a need for alternative technologies that could provide higher efficiencies.

Traditionally, the Shockley-Queisser (SQ) limit [10] has been considered as the maximum theoretical efficiency for single-junction photovoltaic devices at zero bias. The core assumption of SQ is that one photon can generate at maximum one electron-hole pair. Thus, even though hot carriers are generated by high-energy UV photons, their excess energy is assumed to be lost as thermal energy. However, nowadays it is well known that this limit can be overcome by carrier multiplication process, *i.e.* the phenomenon in which the excess energy of hot carriers is utilized to produce further electron-hole pairs by impact ionization [11–13]. Indeed, several examples of internal quantum efficiency (IQE) exceeding one have been reported for both silicon and germanium [14–25] as well as more recently also for graphene/silicon devices [26–28].

While carrier multiplication via high-energy photons would seem an ideal phenomenon to both increase the sensitivity of the sensors and circumvent the SQ limit, to date it has not been successfully demonstrated in real photovoltaic devices. Despite the promising results in IQE, the external quantum efficiencies (EQE) that determine the real device performance, have remained rather modest. There are two fundamental technological obstacles to be tackled in state-of-the-art UV photodiodes [8,9,27–30]: (i) the high reflectance losses and (ii) the extreme sensitivity of photogenerated carriers to surface recombination due to the very shallow absorption depth. Consequently, improvements in the reflectance via micro- and/or nanotexturing are usually counterbalanced by an increase of surface recombination resulting in EQE far from one. Nonetheless, the recently proposed induced-junction black silicon (b-Si) photodiode [31] seems a promising candidate for overcoming the aforementioned obstacles. It consists of a non-reflecting nanostructure combined with efficient surface passivation and junction formation by a charged thin film. The preliminary results have yielded EQE close to 100%. However, the performance in the UV range (below 300nm) has not been confirmed nor the physical phenomena in the induced junction inside nanostructures are known, especially when the UV photons are absorbed only within 10 nm from the surface. Carrier multiplication is likely to take place but, due to the presence of nanostructures, quantum confinement or related phenomena cannot be excluded without further studies.

Here we study the applicability of the induced-junction b-Si photodiodes for UV detection. First, we accurately determine the EQE of the diodes and verify the results by independent measurements at PTB (Physikalisch-Technische Bundesanstalt) with a maximum relative standard uncertainty of 0.4% (see supplemental material, SM [32]). Special emphasis is put on the UV range ($\lambda$=200-350 nm) where we show that extremely high response, more than 130%, can be achieved at zero bias. To gain a more in-depth understanding of the physical phenomena, we first analyze the electric field ($E$-field) and electrostatic potential distributions inside the nanoscale needles using Silvaco Atlas simulations with two typical nano-needle morphologies. Then we examine the IQE rresponding induced-junction photodiode with planar surface. We also compare the features in the IQE spectra to those reported earlier in devices known to possess carrier multiplication. Finally, we simulate the IQE in the nanotextured photodiode and compare that to the experimental one. The results allow us to determine the relevant physical mechanisms leading to an EQE >130%.

Figure 1a shows the EQE and IQE of the b-Si induced-junction photodiode (see inset for schematic device cross-section, Fig. 1 b and c for typical morphologies of the nanostructure, and SM [32] for more detailed description of the device) measured at zero bias in the visible and UV range extending down to 200 nm. The details of the measurements and the certified data can be found in SM [32]. The EQE values exceed 100% at wavelengths shorter than, 310 nm, approximately, and reach values even larger than 130% close to 200 nm. Such high values measured at UV exceed by far all the EQE values reported in literature. In addition, the IQE and the EQE of the device are nearly identical, which is related to the very low reflectance achieved with b-Si (2% av. in the UV). This is a rather unique characteristic clearly demonstrating the superior optical and electronic behavior of the device. The specific features present in the EQE spectrum are discussed later together with the characteristics of the silicon band structure and related band-to-band transitions.

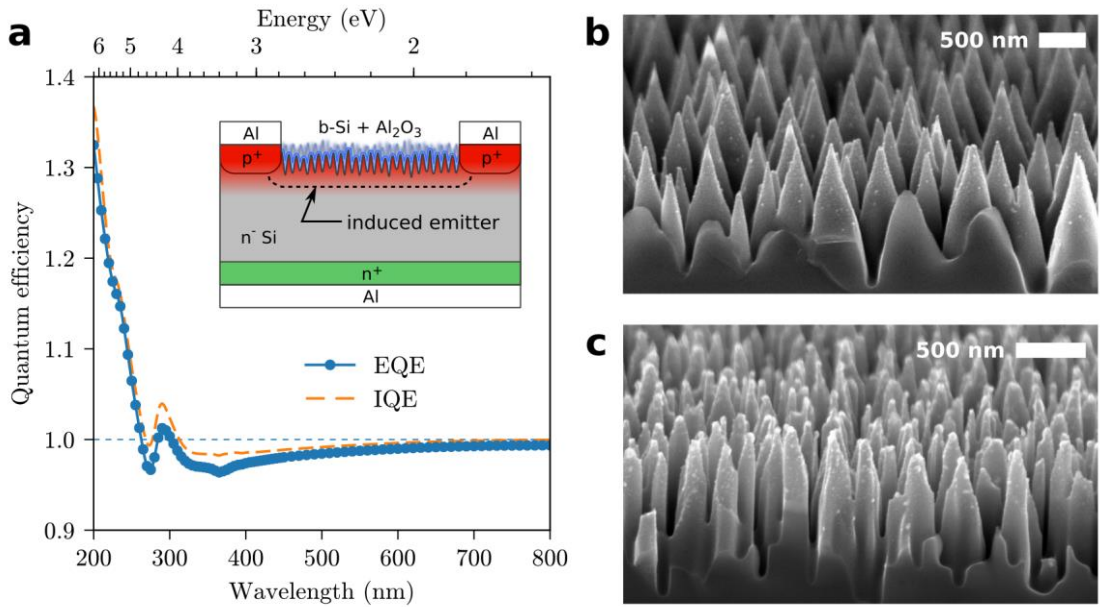

**Fig. 1. a** External (blue circles) and internal (yellow dashed) quantum efficiency of the induced-junction b-Si photodiode measured at zero bias. The inset shows schematically the structure of the device. **b,c** Bird's eye view SEM image of b-Si nanotexture with cone-like and columnar-like morphology, respectively.

The results demonstrate that b-Si induced-junction photodiode structure is particularly sensitive to UV radiation, which is absorbed in the first few nanometers of the device. The excellent sensitivity can be explained as follows: *i*) the induced junction avoids harmful Auger recombination near the front surface, *ii*) a strong *E*-field at the surface ensures immediate collection of the photogenerated carriers, *iii*) nano-texturing efficiently suppresses reflection losses in the UV, and *iv*) surface recombination losses can be minimized due to excellent passivation achieved with $Al_2O_3$. These characteristics alone ensure both collection efficiencies and EQE values close to 100% even without the presence of carrier multiplication or other alternative generation processes. However, in reality at high photon energies carrier multiplication is likely to take place. Therefore, this structure is close to an ideal platform to observe, explore and utilize carrier multiplication.

While it is known that the charge collection efficiency is high in the above device, it is not known whether the presence of ALD coated nanostructures affects the probability of carrier multiplication. In other words, when a photon is absorbed in silicon, whether the b-Si induced junction will boost carrier multiplication in comparison to its planar counterpart with doped *p-n* junction. A possible mechanism for carrier multiplication enhancement could be related to the high charge density present in the $Al_2O_3$, which is typically in the range of $1-5\times10^{12}$ cm$^{-2}$. [41, 42]. The charge induces an intense *E*-field, which is possibly concentrated within the nanostucture, and which should have its maximum near the surface right where the impinging UV photons are absorbed. This could be considered somewhat analogous to avalanche photodiodes, in which a high *E*-field is generated by an external bias voltage. However, here the surface field would most probably assist carrier multiplication only on hot enough carriers resulting from UV photons. To study whether such an effect might be feasible, we show in Fig. 2 both the simulated *E*-field distribution and the electrostatic potential for a single cylindrical-symmetric b-Si nano-needle in thermal equilibrium. We consider the two typical nano-needle morphologies presented in Figure 1 b and c; a cone-like and a columnar-like needle. The simulations were performed with Silvaco Atlas considering a bulk resistivity of 10 kΩ·cm (n-type) and an $Al_2O_3$ fixed charge of $Q_f$=-2.5×10$^{12}$ cm$^{-2}$ (see further details in SM [32]). Notice that this is the typical $Q_f$ value measured for $Al_2O_3$ over b-Si after normalization to the enlarged surface [43]. The simulations reveal that, surprisingly, there is no particular increase in the maximum *E*-field intensity inside the nano-needle, even at the apex of the conic nanostructure. The maximum intensity value stays around 300 kV/cm, which is in the same order as expected for a planar surface with the same level of fixed charge. In general, $E$-fields in the order of $10^6$ V/cm are required to cause a noticeable generation rate by impact ionization in silicon. [44] Although close, the $E$-field induced by the $Al_2O_3$ layer is not high enough to cause impact ionization, especially when taking into account that the field decreases rapidly in just a few nanometers below the surface.

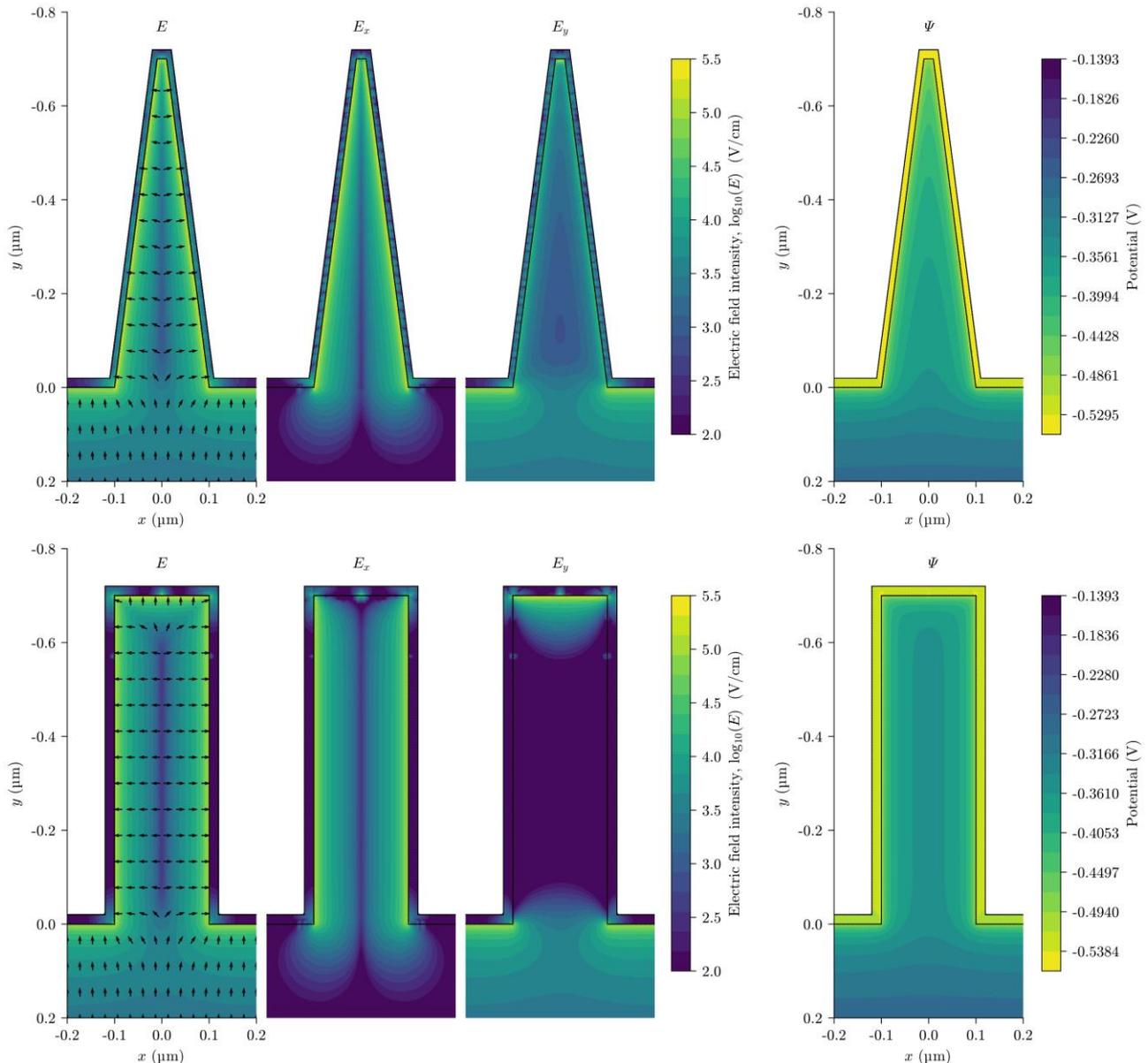

**Fig. 2.** Simulated electrical field distribution ($E$) and electrostatic potential ($\Psi$) in a single b-Si nano-needle with a cone-like shape (top) and a columnar-like shape (bottom).

Although simulations discard any $E$-field hotspots, they reveal an interesting $E$-field distribution inside b-Si. Due to the high aspect ratio of the nano-needles, the $E$-field inside the needle is nearly completely confined into the $x$ direction, this effect being much more pronounced for the columnar geometry. Obviously, $E_x$ neutralizes around the symmetry axis of the needle, but on average $E_y$ stays much lower than $E_x$ throughout the needle volume. On the contrary, $E_y$ noticeably increases at the bottom of the nanostructure resulting in a local maximum, while $E_x$ quickly vanishes outside the needle. Both $E_y$ and $E_x$ simultaneously have low intensity around the symmetry axis of the Si needle due to geometric constraints. Despite of that, the density of holes required to equal the charge $Q_f$ needs to be high in the whole b-Si volume due to both the small dimensions and the large surface-to-volume ratio of the nano-structure. This imposes a high electrostatic potential through the whole volume, including the base of the needle, which then requires a sudden $E$-field increase right at the bottom of the needles. In a sense, it can be understood as if the $y$-component of the $E$-field is shifted down to the bottom of the needle. This particular electrostatic configuration helps masking the nanostructure from the bulk perspective, as the electrostatic potential below the nanostructure quickly flattens at a distance of around 100 nm below the nano-structures.

It is worth to emphasize that the simultaneous combination of low $E$-field and high electrostatic potential ($\Psi$) inside the nanostructure is extremely unusual. In a standard planar device, the presence of $E$-field is always related to the existence of a space-charge-region (SCR) and, conversely, the lack of $E$-field is related to a homogeneously doped neutral region outside the SCR, usually the bulk. In other words, the electrostatic configuration along the core of the b-Si nano-needles is analogous to a homogeneously doped $p$-type region with an equivalent doping density defined by the hole concentration present at the symmetry axis. Notice that the steeper the sidewalls of the needles, the more pronounced this effect is and, as a consequence, in case of vertical cylinders, $E_y$ becomes completely negligible throughout the needle volume (see Fig. 2). The smaller the dimensions of the nanostructure, the higher the equivalent doping would need to be, since the electric dipole at the needle surface becomes confined into a smaller dimension requiring higher induced hole density. Assuming that no other effects arise, this nanoscale "virtual doping" becomes particularly notorious when dimensions are in the order of 10 nm and below, eventually inducing higher $\Psi$ values than expected in the planar counterpart but without increasing the field at the surface. Subsequently, charge carriers generated inside the needles are quickly pushed toward the symmetry axis but, due to the absence of $E_y$-field, their vertical transport is limited by diffusion. The $E_y$ field collects the charge carriers only when they reach the bottom of the nanostructures. Nevertheless, the formation of an induced junction in a nanostructured surface deserves further research in order to fully understand its consequences and dependence on different nanotexture dimensions and geometries. As far as we are concerned in this work, however, the above behavior cannot explain the observed above unity EQE values.

The above simulations strongly suggested that the charged thin film on the nanostructures does not boost carrier multiplication. The next step, therefore, is to compare the measured quantum yield to corresponding planar junction. This should allow us to differentiate between the impact of the nanostructure and the induced junction. Figure 3a shows the IQE and EQE of planar (blue curves) and b-Si (black curves) induced-junction photodiodes, having identical structures besides the surface texture. Somewhat surprisingly, the IQE of the planar photodiode clearly outperforms the b-Si photodiode, reaching IQE values above 200% at a wavelength of 200 nm. For energies above ~ 3.8 eV carrier multiplication becomes clearly visible. For higher energies, the phenomenon increases significantly as a function of photon energy, as predicted by theory. The fact that the phenomenon is clearly more pronounced in planar photodiodes further supports that we can discard any nanoscale-related effects related to b-Si. Furthermore, we can conclude that the effective carrier recombination is higher in b-Si. It is likely that there is a higher amount of recombination sites present in b-Si (reactive ion etching induced damage, larger surface area, exposed crystal planes with varying orientation). Additionally, as the $E$-field is quite small in the middle of the needles (Fig.2), the carriers need to diffuse a relatively long distance before being collected which may also increase recombination. This leads to the possibility that by reducing the aspect ratio without compromising the reflectance, the performance of b-Si photodiodes could be even higher. Obviously, other considerations, such as the enlarged effective surface and a greater density of states in the textured surface, need to be factored in during the optimization. This result also suggests that there is plenty of room for improvement in planar UV photodiodes by incorporating an induced junction in combination with traditional antireflection coatings.

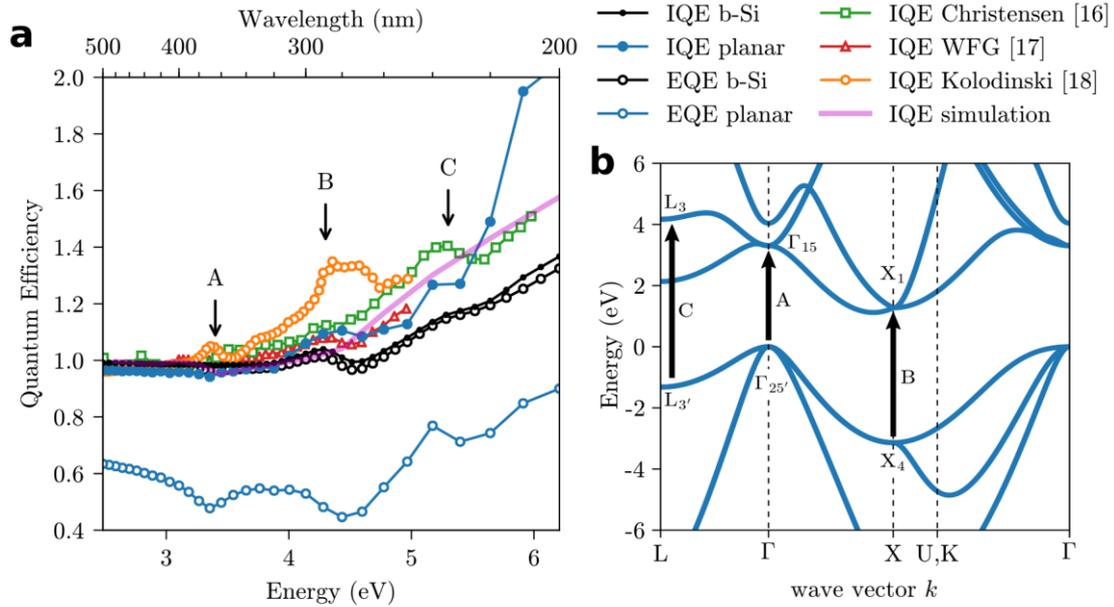

**Fig. 3. a** Comparison of both IQE and EQE of our two similar induced junction photodiodes, one with b-Si texturing (black curves) and another with planar surface (blue curves). Additionally, the IQEs reported for different silicon photodetectors with planar surface are shown as comparison. The purple curve shows our Silvaco Atlas simulation of b-Si photodiode. **b** Schematic of the band diagram for silicon.

Further insight on the physical nature of carrier multiplication in our photodiodes can be achieved through carefully inspecting the features of the IQE spectra in the UV range. Figure 3a shows the comparison to previously published diodes [16–18] that possess high IQE values, above one, due to carrier multiplication. All the curves show similar spectral features that can be related to characteristics in the band structure of silicon (see Figure 3b). It should be pointed out that in this energy range carrier multiplication is particularly sensitive to the band structure and the related wave vector location. More specifically, the probability of carrier multiplication peaks at the direct transition points such as $\Gamma$ point [18] followed by other high symmetry points (X and L). In Figure 3a, the most prominent features visible in the IQE curves are labelled as B and C. Feature B, present at all curves as a peak at around 4.3 eV, can be associated to impact ionization by holes [18]. It corresponds to the transition $X_4$-$X_1$ ($E_2$ transition), which generates hot holes that would be responsible for the carrier generation by impact ionization. The spectral feature C, at around 5.3 eV, a peak in Christensen's data but a shoulder in our measurements, is likely related to the direct transition $L_{3'}$-$L_3$ ($E_1'$ transition) [45]. Interestingly, feature A is non-existent in our measurement in both b-Si and planar samples. It should appear at around 3.4 eV and has been associated to impact ionization by hot electrons generated through the transition $\Gamma_{25'}$-$\Gamma_{15}$ ($E_0'$ transition) by Kolodinski *et al.* [18] The explanation that we do not see the feature A could be related to base doping (n-type) combined with induced junction (as compared to *p*-type doping with Kolodinski) and thus the probability for hot electrons would be reduced. However, we cannot totally rule out the role of inaccuracies in reflectance measurements, which are usually relatively high for wavelengths below 350 nm, since also the dispersion in silicon show features in the UV related to the band structure. In summary, all the observations in b-Si are similar to the results obtained in planar devices thereby discarding *e.g.* quantum-confinement effects in the b-Si nanostructure tips.

In order to support the experimentally observed above unity EQE and the proposed explanation for it, we performed simulations of the device IQE using Silvaco Atlas and a model of the b-Si photodetector described in detail in Ref. [45]. Since Silvaco Atlas does not incorporate multiple carrier generation in their models, this needs to be separately accounted for in the simulations. We did this by considering the calculations presented in Ref. [13] for the mean energy required for creating an electron-hole pair as a function of the photon energy and by adjusting the number of carriers generated by each photon at each wavelength, correspondingly. This should result in a generation profile resembling an experimental one. (Further details of the simulations are given in SM [32]). The purple line in Fig. 3a presents the simulated IQE in the b-Si detector. The shape matches nicely the experimental data points, excluding the small features since the Ref. [13] does not take into account the specific transitions (A, B and C) in the band structure. The slightly higher IQE above 4.5 eV in simulations can be explained by near-surface etching damage causing additional recombination in the experimental IQE. All in all, the simulations support the measured above unity EQE.

We have shown that EQE values over 100% in b-Si induced-junction devices can be achieved in a broad spectral range in the UV. This result is significant, as it has been achieved without applying any extra energy, such as bias voltage. Moreover, the much higher IQE observed in planar devices strongly suggests that there might be still room for further improvement in both texturized devices (through texture optimization) and conventional planar devices. In principle, since nanotexturization does not seem to directly boost the carrier multiplication, the low reflectance could also be achieved by conventional antireflection coating (ARC). However, the nanotexturing route has several benefits in comparison to ARC, including wide acceptance angle and extremely low reflectance over a wide spectral range [46]. While there might be some concerns about the industrial applicability of the nanostructures, it has been shown that the b-Si nanostructures survive well in the industrial mass production lines [47] and their manufacturing costs can be competitive even in the cost-driven photovoltaics industry [48]. Furthermore, there are also alternative cheaper nanotexturing technologies available such as metal assisted chemical etching that can be considered [49,50].

While the current results can readily have a high impact on the UV-sensor industry, they are promising considering other applications as well, namely, above unity responses could be possible for longer wavelengths too. For instance by applying similar methods, *i.e.* nanostructures and induced junction in lower band gap materials (such as germanium), carrier multiplication should be present and, if efficiently harvested, the response and gain in the visible part of the spectrum would be enhanced. Another important consequence is that the thermal losses can be reduced considerably, which minimizes thermal noise and the need for cooling in photovoltaic devices. In addition to photodiodes, solar cells could benefit from carrier multiplication if lower bandgap material could be used to efficiently capture the total energy of the photons without suffering from thermal losses. This applies especially to space applications where the intensity of UV radiation is known to be higher. Finally, the present work brings us a step closer to the long sought goal of surpassing the SQ efficiency limit, offering new avenues for an effective utilization of carrier multiplication effect in semiconductors.

In conclusion, we reported certified measurements that demonstrate EQE above unity in a b-Si induced-junction photodiode without external amplification. In particular, we showed EQE values rising up to 132% at 200 nm wavelength. The effect of b-Si was investigated through: i) numerical simulations of the *E*-field and electrostatic potential in the nanostructure, ii) comparison of IQE to the corresponding planar photodiode, iii) analysis of features in the IQE spectra, and iv) simulations of IQE in b-Si photodiode. All the results consistently showed that the high performance is based on effective utilization of multiple carrier generation by impact ionization taking place in the nanostructures. The results suggest that utilizing similar concept in lower band gap materials should be relatively straightforward, possibly extending the above unity performance to lower photon energies.

Acknowledgements: The authors acknowledge the provision of facilities and technical support for sample fabrication and characterization by Micronova Nanofabrication Centre in Espoo, Finland, within the OtaNano research infrastructure at Aalto University. T. P. Pasanen, V. Vähänissi and H. Savin acknowledge the funding from the ATTRACT project funded by the European Council (EC) under Grant Agreement 777222 and the funding from the Academy of Finland. The work is related to the Flagship on Photonics Research and Innovation "PREIN" funded by the Academy of Finland. M. G. acknowledges support from the project ENE2015-74009-JIN of the Spanish Ministry of Economy and Competitiveness (MINECO) and co-funded by the European Regional Development Fund. J. Heinonen, A. Haarahiltunen and M. Juntunen acknowledge financial support from Business Finland.